    \documentstyle[epsf,fancyhdr,pilscap,twoside]{article} \catcode`\@=11
    \def\section{\@startsection{section}{1}{\z@}%
    {-5.5ex plus -1ex minus -.5ex}{1.5ex plus.3ex}{\bf }}
    \def\subsection{\@startsection{subsection}{1}{\z@}%
    {-3.5ex plus-1ex minus-.5ex}{1.5ex plus.3ex}{\sl }} 
\begin{document}
   \setcounter{page}{493}
    \lhead{Ann. Phys. (Leipzig) {\bf 7} (1998) 5---6, 493---497}
    \cfoot{}
    \vspace*{2.1cm} 
{\Large\bf
Electronic transport \\[.2\baselineskip]
in disordered interacting systems
    }\vspace{.4cm}\newline{\bf   
 Thomas Vojta and Frank Epperlein
    }\vspace{.4cm}\newline\small
  Institut f{\"u}r Physik, Technische Universit{\"a}t, D-09107
  Chemnitz, Germany
    \vspace{.2cm}\newline 
Received 6 October 1998, accepted 8 October 1998 by B. Kramer
    \vspace{.4cm}\newline\begin{minipage}[h]{\textwidth}\baselineskip=10pt
    {\bf  Abstract.}
We numerically investigate the transport properties of disordered interacting 
electrons  in three dimensions in the metallic as well
as in the insulating phases. The disordered many-particle problem is modeled by the
quantum Coulomb glass which contains  a random potential, long-range unscreened
Coulomb interactions and quantum hopping between different sites.
We have recently developed the Hartree-Fock based diagonalization (HFD) method
which amounts to diagonalizing the 
Hamiltonian in a suitably chosen energetically 
truncated basis. This method allows us to investigate comparatively large systems.
Here we calculate the combined effect of disorder and 
interactions on the dissipative conductance. We find that the qualitative
influence of the interactions on the conductance depends on
the relative disorder strength. For strong disorder interactions can significantly enhance 
the transport while they suppress the conductance for weak disorder. 

    \end{minipage}\vspace{.4cm} \newline {\bf  Keywords:}
Disorder; Anderson localization; Interactions; Correlated electrons
    \newline\vspace{-.15cm} \normalsize
\section{Introduction}
After four decades of research the transport properties of systems in which both disorder and
strong interactions are equally important are still not even qualitatively understood. 
Based on Andersons seminal paper \cite{anderson}
investigations of {\em non-interacting} disordered electrons led to the development 
of the scaling theory of localization
\cite{aalr,kramrev93}. In the absence of external symmetry breaking it predicts
that all states are localized in one and two dimensions. In contrast, in three dimensions states 
are extended for weak disorder while they are localized for sufficiently strong disorder. 
This gives rise to a disorder-driven metal-insulator transition (MIT) at a certain value of disorder strength.

Later also the influence of electron-electron interactions on the transport properties of disordered
electrons was investigated intensively by means of  many-body perturbation theory, field theory, and
the renormalization group \cite{leerev85,polrev85,belitzrev94}.  This led to a qualitative analysis of 
the MIT and the identification of the different universality classes. 
The topic has reattracted a lot of attention after unexpected experimental \cite{2DMIT} and 
theoretical \cite{TIP} findings.
In order to investigate the localized phase and to check the validity of the perturbative results
it seems to be important to investigate the problem of interacting disordered
electrons non-perturbatively. One possible way in this direction is numerical simulations
although they are very costly for disordered many-body systems.  
Recently, we have developed  the Hartree-Fock based diagonalization (HFD) method \cite{HFD,method}
for the simulation of disordered quantum many-particle systems.
We have already applied this very efficient method for calculating the transport properties
of one-dimensional \cite{giessen} and two-dimensional \cite{HFD} disordered interacting electrons. 
In this paper we extend the investigation to three spatial dimensions.

\section{Model and calculations}
\fancyhead[LE,RO]{\thepage}
\fancyhead[LO]{T. Vojta, F. Epperlein: Transport in disordered interacting systems}
\fancyhead[RE]{Ann. Phys. (Leipzig) {\bf 7} (1998) 5---6}

The model, called the quantum Coulomb glass model 
\cite{efros95,talamantes96,epper_hf,epper_exact},  is defined on a cubic lattice of $L^3$ 
sites occupied by $N=K L^3 $ electrons ($0\!<\!K\!<\!1$). To ensure charge neutrality
each lattice site carries a compensating positive charge of  $Ke$. The Hamiltonian
is given by
\begin{equation}
H =  -t  \sum_{\langle ij\rangle} (c_i^\dagger c_j + c_j^\dagger c_i) +
       \sum_i \varphi_i  n_i + \frac{1}{2}\sum_{i\not=j}(n_i-K)(n_j-K)U_{ij}
\label{eq:Hamiltonian}
\end{equation}
where $c_i^\dagger$ and $c_i$ are the electron creation and annihilation operators
at site $i$, respectively,  and $\langle ij \rangle$ denotes all pairs of nearest 
neighbor sites.
$t$ gives the strength of the hopping term and $n_i$ is the occupation number of site $i$. 
For a correct description of the insulating phase the Coulomb 
interaction between the electrons is kept long-ranged,
$U_{ij} = U/r_{ij}$ ($r_{ij}$ is measured in units of the lattice constant), 
since screening breaks down in the insulator. 
The random potential values $\varphi_i$ are chosen 
independently from a box distribution of width $2 W$ and zero mean.
(In the following we always set $W=1$.)
Two important limiting cases of the quantum Coulomb glass are the Anderson model of
localization (for $U_{ij}=0$) and the classical Coulomb glass (for $t=0$).
For the present study we have simulated systems with $3^3$ sites and 13 electrons
using periodic boundary conditions.

The calculations are carried out by means of the  HFD method \cite{HFD}.
This method which is based on the idea of the configuration interaction approach
\cite{fulde} adapted to disordered lattice models is
very efficient in calculating low-energy properties in any spatial dimension and for 
short-range as well as long-range interactions.
It consists of 3 steps: (i) solve the Hartree-Fock (HF) approximation of the Hamiltonian, 
(ii) use a Monte-Carlo algorithm to find the low-energy many-particle HF states, 
(iii) diagonalize the Hamiltonian in the basis formed by these states. 
The efficiency of the HFD method is due to the fact that the HF basis states are 
comparatively close in character to the exact eigenstates in the entire
parameter space.
Thus it is sufficient to keep only a small fraction of the Hilbert 
space in order to obtain low-energy  quantities with an accuracy comparable to that of exact 
diagonalization. 
For the present systems of 13 electrons on 27 sites  we found it sufficient to keep 500 to 1000  
(out of $2*10^8$) basis states.

In order to calculate the conductance we employ the Kubo-Greenwood 
formula \cite{kubo_greenwood} which connects the conductance with the
current-current correlation function in the ground state. Using the spectral
representation of the correlation function the real (dissipative) part 
of the conductance (in units of the conductance quantum $e^2/h$) is obtained as 
\begin{equation}
 Re ~ G^{xx}(\omega) = \frac {2 \pi^2 L}   {\omega} \sum_{\nu} |\langle 0 | j^x|\nu \rangle |^2 
     \delta(\omega+E_0-E_{\nu})
\label{eq:kubo}
\end{equation}
where $j^x$ is the $x$ component of the current operator and $\nu$ denotes the eigenstates
of the Hamiltonian.  The finite life time $\tau$ of the eigenstates in a real d.c.\ transport experiment
(where the system is not isolated but connected 
to contacts and leads) results in an inhomogeneous broadening $\gamma = 1/\tau$
of the $\delta$ functions in the Kubo-Greenwood formula \cite{datta}. Here we use $\gamma=0.05$
which is of the order of the single-particle level spacing. Since the resulting
conductances vary strongly from sample to sample we logarithmically average all results 
over several disorder configurations.
\begin{figure}[th] 
  \epsfxsize=10.5cm
  \centerline{\epsffile{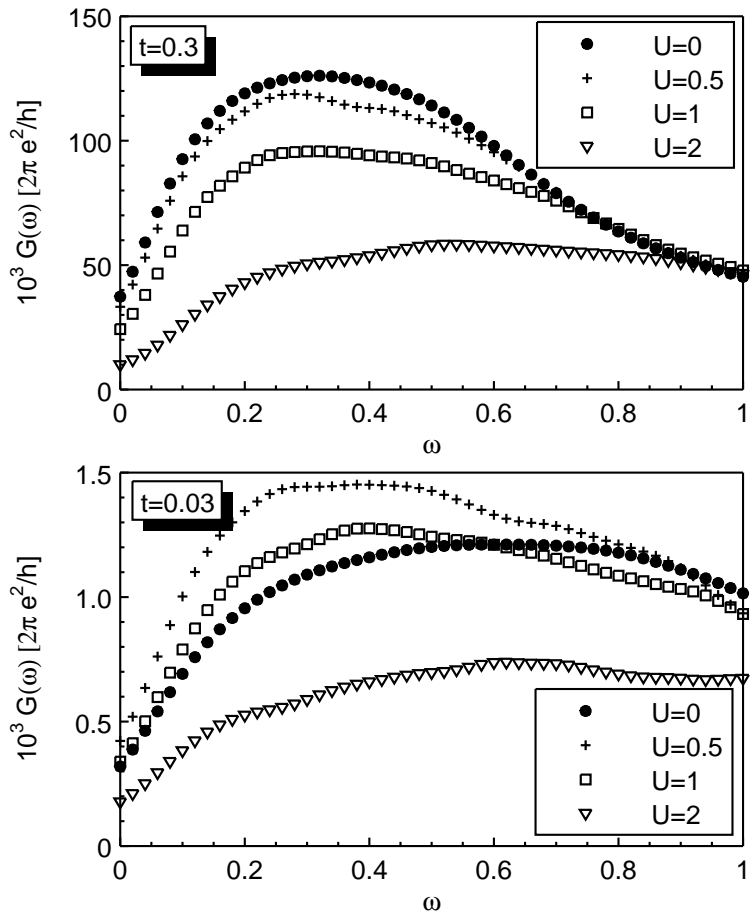}}
  \caption{\protect\small Conductance $G$ as a function of frequency $\omega$,
              $W=1$,$\gamma=0.05$ and two different values of the kinetic energy $t$.  
              The truncation of the Hilbert space
              to 500 basis states restricts the validity of these data to frequencies $\omega<0.75$.}
  \label{fig:sig}
\end{figure}

\section{Results}
\begin{figure}[th]
  \epsfxsize=10.5cm
  \centerline{\epsffile{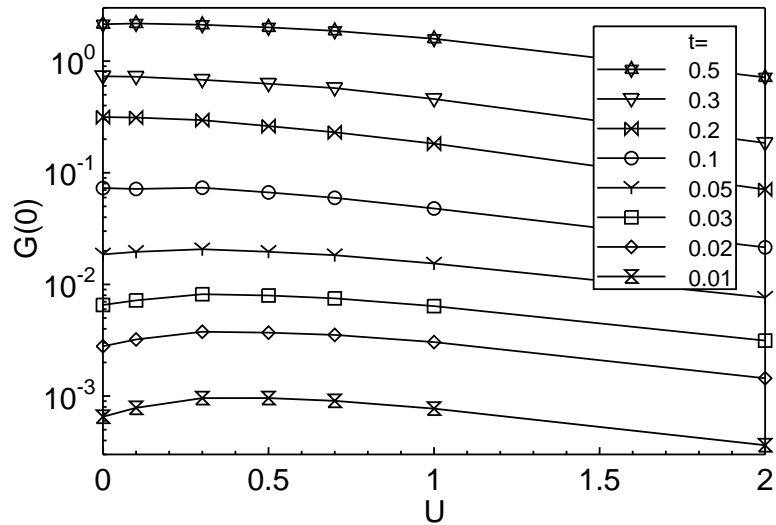}}
  \caption{\protect\small d.c.\ conductance $G(0)$ as a function of interaction strength $U$
              for different kinetic energies $t$.}
  \label{fig:sigzero}
\end{figure}

We now present results on the dependence of the conductance on the
interaction strength.
In Fig.\ \ref{fig:sig}  we show the conductance as a function 
of frequency  for two sets of parameters. The data represent logarithmic averages 
over 400 disorder  configurations.
In the lower diagram the kinetic energy is very small ($t=0.03$), i.e., the system is in the 
highly localized regime.
Here not too strong Coulomb interactions ($U = 0.5, 1.0$) lead to
an {\em increase} of the conductance at low frequencies. If the interaction 
becomes stronger ($U=2$) the conductance finally decreases again.
The behavior is qualitatively different at higher kinetic energy 
($t=0.3$) as shown in the upper diagram. Here
already  a weak interaction ($U=0.5$) leads to a reduction of the 
low-frequency conductance
compared to non-interacting electrons. If the interaction becomes stronger ($U=2$)
the conductance decreases further.
We have carried out analogous calculations for kinetic energies $t=0.01, ..., 0.5$ and
interaction strengths $U=0, ..., 2$. The resulting d.c.\ conductances are presented
 in Fig.\ \ref{fig:sigzero} which is the main result of 
this paper.
In Fig.\ \ref{fig:sigzero} we now present results on the dependence 
of the d.c.\ conductance $G(0)$ on the
interaction strength for kinetic energies $t=0.01, ..., 0.5$ and
interaction strength $U=0, ..., 2$. 
The data  show that the influence of weak repulsive electron-electron interactions on the d.c.\ 
conductance depends on the ratio between disorder and kinetic energy.
The conductance of 
strongly localized samples ($t=0.01, ..., 0.03$) becomes considerably enhanced 
by a weak Coulomb interaction which can be understood from the suppression of localizing
interferences by electron-electron scattering. With increasing kinetic energy the relative enhancement
decreases as does the interaction range where the enhancement occurs. The conductance
of samples with high kinetic energies ($t \ge 0.1$) is not enhanced  
by weak interactions.  
Sufficiently strong interactions always reduce the conductance. In this regime the
main effect of the interactions is the reduction of charge fluctuations which reduces
the conductance. In the limit of  infinite interaction strength the system approaches a 
Wigner crystal/glass which is insulating for any disorder.
Overall, this is the same qualitative 
behavior as in the cases of one \cite{giessen} and two \cite{HFD} spatial dimensions
although the disorder induced enhancement of the conductance seems to be weaker in three 
dimensions. Up to now it is not clear whether this is a true dimensionality effect or due
to the different linear system sizes studied. A systematic investigation of the size
dependence is in progress.

To summarize, we have used the Hartree-Fock based diagonalization (HFD) method
to investigate the conductance and localization properties of disordered interacting spinless
electrons in three dimensions. We have found that a weak Coulomb interaction 
can enhance the conductivity of strongly localized samples by almost one order of magnitude,
while it reduces the conductance of weakly disordered samples. If the interaction becomes stronger 
it eventually reduces the conductance also in the strongly localized regime. 
\\  

\baselineskip=10pt{\small 
We acknowledge financial support by the Deutsche Forschungsgemeinschaft.
    }
    \end{document}